# The Structure of the Solar Core

**O Manuel**
University of Missouri, Nuclear Chemistry, Rolla, Missouri 65401, U.S.A.

**Abstract.** The Apollo and Galileo missions provided unambiguous evidence that the Sun is iron-rich. New experimental measurements and theoretical studies are now needed to build a framework for understanding the source of solar luminosity, neutrinos and mono-isotopic hydrogen coming from the Sun. Specifically, measurements are needed of the low energy (E < 0.782 MeV) anti-neutrinos that would be produced if neutron-decay is occurring near the core of the Sun and a theoretical basis is needed to understand empirical evidence of a) repulsive interactions between like nucleons, b) clustering of nucleons, and c) possible neutron penetration of the gravitational barrier surrounding a neutron star.

## 1. Introduction

The Apollo program landed men on the moon six times from 1969 to 1972 and returned lunar samples for laboratory analysis at a total cost of $150 billion to $175 billion US dollars [1]. This "cold-war" mission provided unexpected new information about the Sun. Material implanted in the surfaces of lunar samples by the solar wind (SW) revealed unambiguous evidence of severe mass fractionation (MF) that enriches lighter mass (L) elements and isotopes over heavier mass (H) ones at the solar surface by a common power law [2], where the mass fractionation is

$$MF = (H/L)^{4.56} \qquad (1)$$

When the elemental composition of the solar photosphere[3] is corrected for this empirical fractionation, the most abundant elements in the interior of the Sun are found to be Fe, Ni, O, Si, S, Mg and Ca [2]. These same seven, even-Z elements comprise 99% of the material in ordinary meteorites [4]. The probability (P) is essentially zero that this agreement is fortuitous, P < 0.00000000000000000000000000002.

The Galileo mission that reached Jupiter in 1996 was less expensive but equally important in confirming the prediction [ref. 2, p. 220] of "strange" xenon in Jupiter, unlike that in the Sun. The data are available on the web at http://web.umr.edu/~om/abstracts2001/windleranalysis.pdf

Thus, the Apollo and Galileo missions confirmed that the interior of the Sun is iron-rich, as indicated by numerous earlier analyses of meteorites [5,6,7], terrestrial planets [8], and the Sun itself [9,10]. Ref. [11] has a concise list of the measurements since 1960, which indicate that the Sun formed on the collapsed core of a supernova, as shown in Fig. 1.

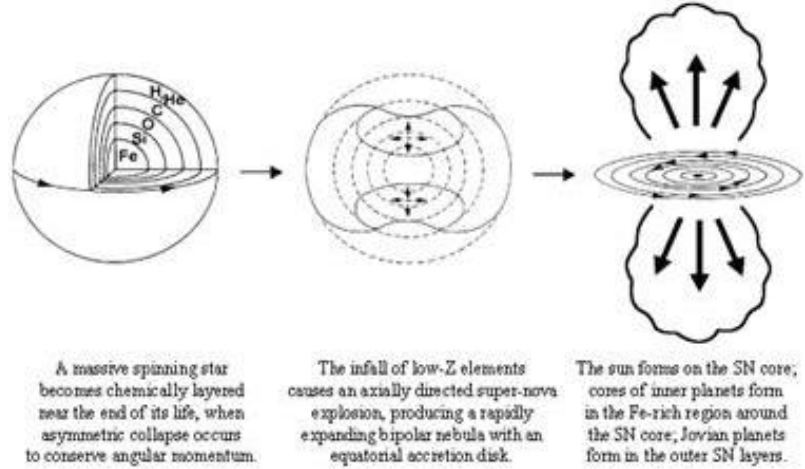

**Figure 1.** The Solar System formed from debris of a spinning supernova.

The last frame in Fig. 1 is remarkably like a recent photo from the Hubble space telescope of a nearby dying star called V Hydrae: http://www.universetoday.com/am/publish/v_hydra_finale.html This discovery was the subject of a recent report in *Nature* [12], but the pictures published there are less clear. Proponents of the hydrogen-filled Sun should now address the measurements listed in ref. [11] and others that indicate the Sun is iron-rich [5-10] and oscillates like a pulsar [13].

At last year's conference in Oulu, Finland [14] it was noted that neutron emission from the central neutron star at the core of the Sun likely triggers a series of reactions that generate solar luminosity (SL), neutrinos, and a upward flow of protons that maintains mass separation in the Sun and generates an outpouring of $H^+$ ions from the solar surface:

- Neutron emission from a central neutron star ( >57% SL)
$$<{}^1_0n> \longrightarrow {}^1_0n + \sim 10\text{-}22 \text{ MeV}$$
- Neutron decay ( <5% SL)
$$ {}^1_0n \longrightarrow {}^1_1H^+ + e^- + \textbf{anti-n} + 0.782 \text{ MeV}$$
- Fusion and upward migration of $H^+$ ( <38% SL)
$$4\, {}^1_1H^+ + 2\,e^- \longrightarrow {}^4_2He^{++} + 2\,\textbf{n} + 27 \text{ MeV}$$
- Escape of excess $H^+$ in the solar wind (100% SW)
Each year $3 \times 10^{43}$ $H^+$ depart in the solar wind.

Abrupt changes in climate and the heterogeneous, dynamic nature of the Sun have also been at odds with the assumption of a homogeneous Sun with a well-behaved H-fusion reactor at its core. Many of these violent events at the solar surface are driven by solar magnetic fields, deep-seated remnants of ancient origin [15] arising from a) the neutron star at the solar core, and/or b) the iron-rich, super-conducting [16] material that surrounds the central neutron star.

The present paper identifies the need for a better theoretical under-standing of the processes that occur in an iron-rich Sun and suggests a few experimental measurements to test if these are part of the Sun's operation.

## 2. The source of luminosity in an iron-rich Sun

Over 20 years ago it became abundantly clear that the Sun must be iron-rich [2]. However the stable isotopes of iron contain tightly bound nucleons, so this could hardly be the source of solar luminosity. Finally on Christmas day of 2000, three students and I submitted a report to the Foundation for Chemical Research, Inc. [17] with a summary of information obtained when we abandoned the conventional approach and used something akin to the reduced variables in van der Waals' equation of corresponding states to study properties of the 2,850 nuclides tabulated in the latest report from the National Nuclear Data Center [18].

Trends in the reduced variables, Z/A or charge per nucleon, and M/A or potential energy per nucleon, revealed evidence that the n-p interactions are strongly attractive, while the n-n and p-p interactions are strongly repulsive and symmetric after correcting for the well-known repulsive Coulomb interactions between positive nuclear charges.

These findings were unexpected from the two-nucleon interaction potentials described in standard nuclear textbooks, where disintegration, far from the valley of beta stability, is attributed to proton and neutron drip lines beyond which "… *the unbound proton or neutron drips out of the nucleus.*" [19, page 381].

On the contrary, trends in the empirical data indicate that neutron or proton emission releases large amounts of energy if the parent nuclide is far from the valley of beta stability. Proton emission releases the largest amount of energy when Z/A ~ 1.0, but this probably does not correspond to any natural form of matter heavier than $^1$H. However, neutron emission when the parent nuclide has Z/A ~ 0, e.g., a neutron star, typically releases 10-22 MeV per neutron emitted. This converts a larger fraction of rest mass into energy than does fission or fusion. Thus, neutron emission may account for a large fraction of the energy released by the Sun and other stars that formed on collapsed supernova cores [5, 6].

The empirical basis for concluding that interactions between like nucleons are repulsive, that nucleons cluster, and that neutron emission may be a major source of solar energy is presented below. These empirical findings and their possible roles in the operation of the Sun demonstrate the need for a better theoretical understanding of nucleon interaction potentials, and possible neutron penetration of the gravitational potential barrier around a neutron star. The results may advance both nuclear physics and our understanding of the source of energy that bathes planet Earth and sustains life.

### 3. Nuclear clustering and interactions between nucleons

The "cradle of the nuclides", Fig. 2, illustrates major trends when data for ground states of the 2,850 known nuclides [18] are plotted in terms of Z/A, charge per nucleon, versus M/A, mass or total potential energy per nucleon, and then sorted by mass number, A. All nuclides have values of $0 = Z/A = 1$, and these define a cradle shaped like a trough made by holding two cupped hands together. The more stable nuclides lie along the valley, and $^{56}$Fe lies at the lowest point. Lighter, more fusible nuclides occupy higher positions, up the steep slope to the left of A = 56 in Fig. 2. The heavier, more fissionable nuclides occupy slightly higher positions, up the gradual slope to the right of A = 56.

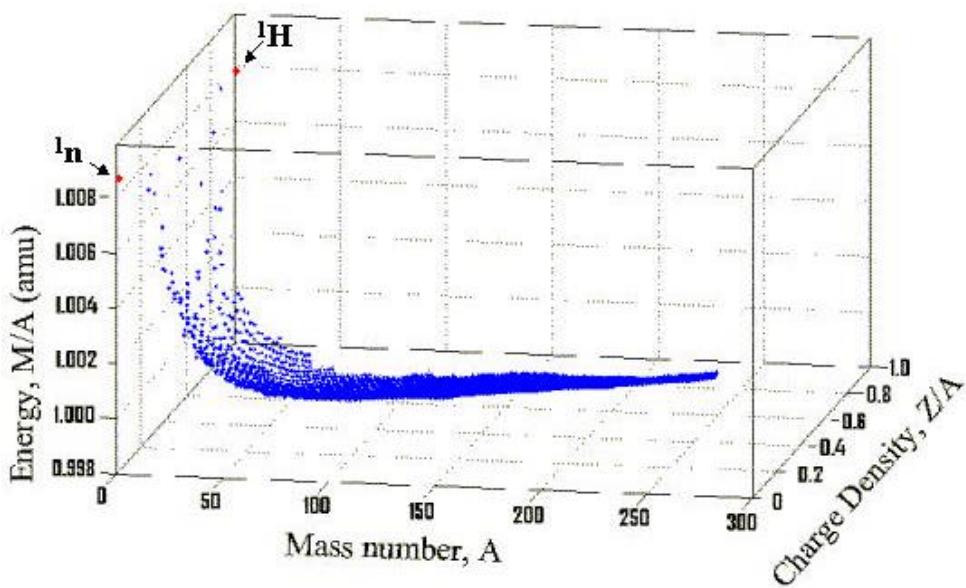

**Figure 2**. The ground states of the 2,850 nuclides define a "cradle".

At any given value of A, the masses of the nuclides define a "mass parabola" as the values of Z/A increase from the lowest known value, closest to Z/A = 0, to the highest known value, closest to Z/A = 1. The most stable charge on any nuclide of mass number A generally lies about midway between the front and back planes in Fig. 2, at the low point in the mass parabola.

Nuclides that are closer to the front plane, i.e., those having lower values of Z/A, tend to decay by negatron (electron) emission; nuclides that are closer to the back plane in Fig. 2, i.e., those having higher values of Z/A, tend to decay by positron emission or electron capture. There is a minor "saw-tooth" fine structure caused by even-even versus odd-odd effects when A is an even number. To avoid this distraction, the next three graphs will show these trends in more detail when A is an odd number.

Fig. 3 shows, for example, a cross section through Fig. 2 at A = 27. The low point in the mass parabola occurs at $^{27}$Al. From left to right, all eight known nuclides [18] at A = 27 are $^{27}$F, $^{27}$Ne, $^{27}$Na, $^{27}$Mg, $^{27}$Al, $^{27}$Si, $^{27}$P, and $^{27}$S.

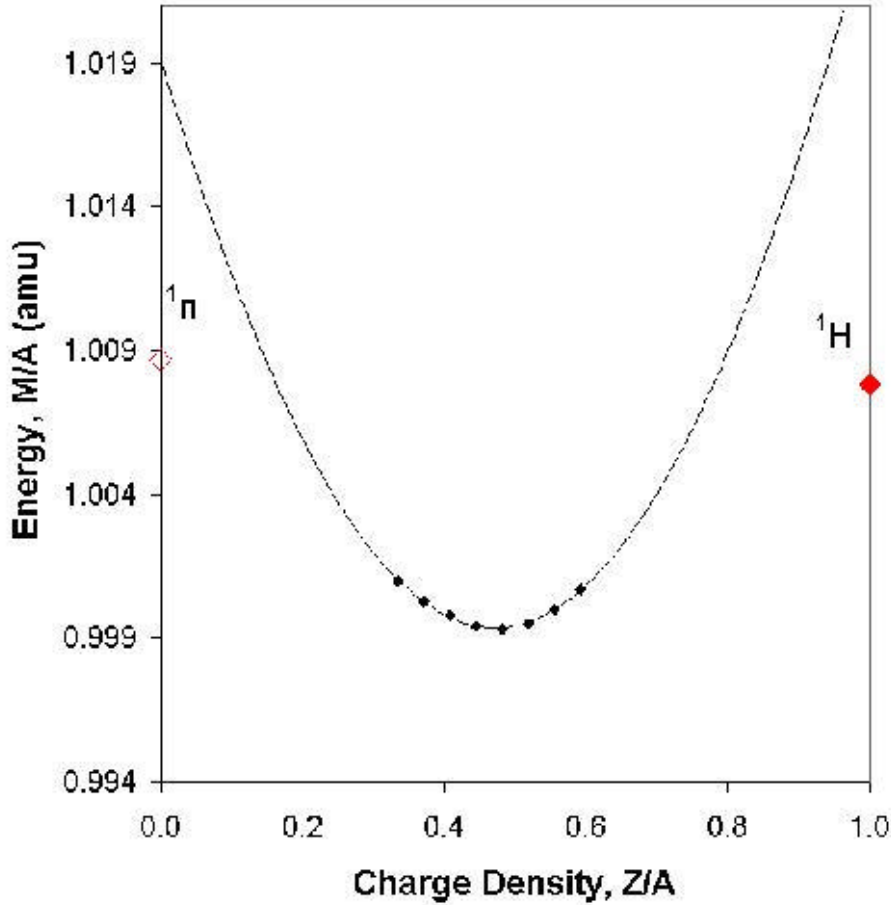

**Figure 3.** An illustrative cross section through Fig. 2 at A = 27.

Fig. 3 also shows values of M/A for unbound nucleons, a neutron on the left at Z/A = 0 and an $^1$H atom on the right Z/A = 1.0, respectively. The empirical mass parabola defined by $^{27}$F, $^{27}$Ne, $^{27}$Na, $^{27}$Mg, $^{27}$Al, $^{27}$Si, $^{27}$P, and $^{27}$S yields much higher values of M/A for an assemblage of 27 neutrons at Z/A = 0 or for an assemblage of 27 protons at Z/A = 1.0, respectively. Cross-sectional cuts through Fig. 2 at any other value of A > 1 reveal an empirical mass parabola with values of M/A > M($^1$n) at Z/A = 0 and values of M/A > M($^1$H) at Z/A = 1.0.

Typically the excess energy associated with these assemblages of pure neutrons or protons is ~10 MeV per nucleon, plus the energy from Coulomb repulsion at Z/A = 1. Unlike the imagined dripping of neutrons near Z/A ~ 0 [ref. 19, page 381], it thus appears that neutron emission may release significant amounts of energy from a neutron star.

Coulomb repulsion contributes to the high value of M/A for the assemblage of 27 protons on the right side of Fig. 3, but not to a nucleus of 27 neutrons on the left. In fact, Coulomb repulsion accounts for the difference between values of M/A at the intercepts where Z/A = 1.0 and Z/A = 0, and this difference increases linearly with $A^{2/3}$ over the mass range, A = 1 - 41 [20]. The slope of this line is indistinguishable from that defined by the familiar β-decay of mirror nuclei close to the line of β-stability, e.g., ($^1$H, $^1$n), ($^3$He, $^3$H), ($^5$Li, $^5$He), ($^7$Be, $^7$Li,), . . . , ($^{41}$Sc, $^{41}$Ca) [20].

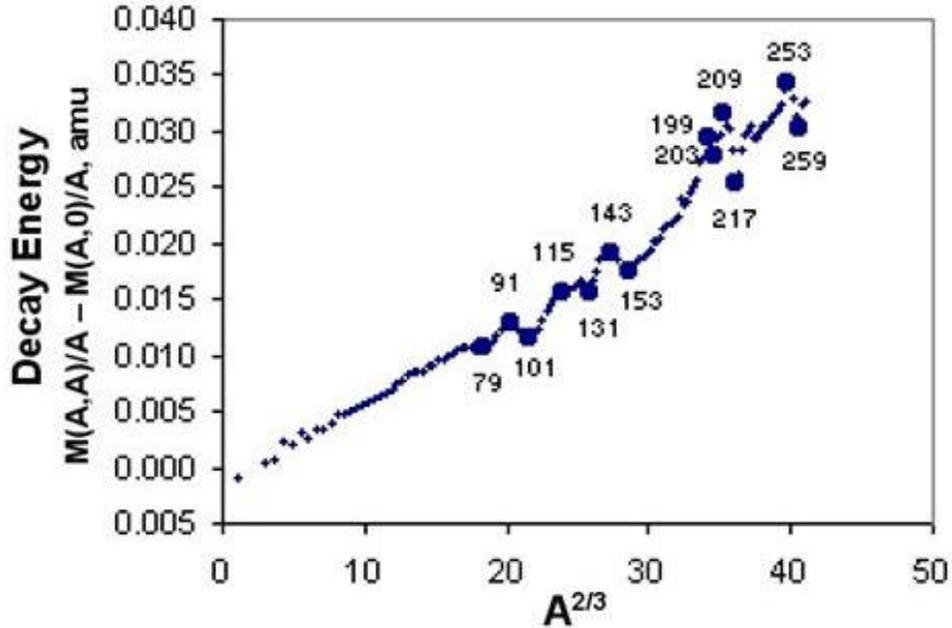

**Figure 4**. Decay energies of extreme nuclides, where the Coulomb energy drives (Z/A = 1) —> (Z/A = 0), for all odd values of A from A = 1 to 263.

Thus, the values obtained for M/A from empirical mass parabolas at Z/A = 1.0 and Z/A = 0 yield the same nuclear radius and the same coefficient for the Coulomb energy term as the mirror nuclei close to the line of β- stability for A = 1 - 41 [20].

The decay energy, and hence the Coulomb energy of heavier nuclides, A > 41, can also be obtained from differences indicated by mass parabolas for values of M/A at Z/A = 1.0 and Z/A = 0. Fig. 4 shows the results for all odd values of A, from A = 1 to A = 263.

The decay energies of light nuclides in Fig. 4 vary linearly with $A^{2/3}$, but fine structure starts to appear near A ~ 80. Peak energies, at A = 91, 115, 143, 199, 209 and 253, likely arise from high Coulomb energy at Z/A = 1 because of clustering of nucleons into tightly packed structures. Likewise, valleys at A = 79, 101, 131, 153, 203, 217 and 259 likely mean low Coulomb energy at Z/A = 1 because of more loosely packed nucleons.

There is no Coulomb energy associated with the other extreme form of nuclides, at Z/A = 0. These are the intercepts of mass parabolas at each value of A with the front plane in Fig. 2. However, these neutron-rich nuclides at Z/A = 0 also reveal fine structure, as shown in Fig. 5 for all odd values of A from A = 1 to 263.

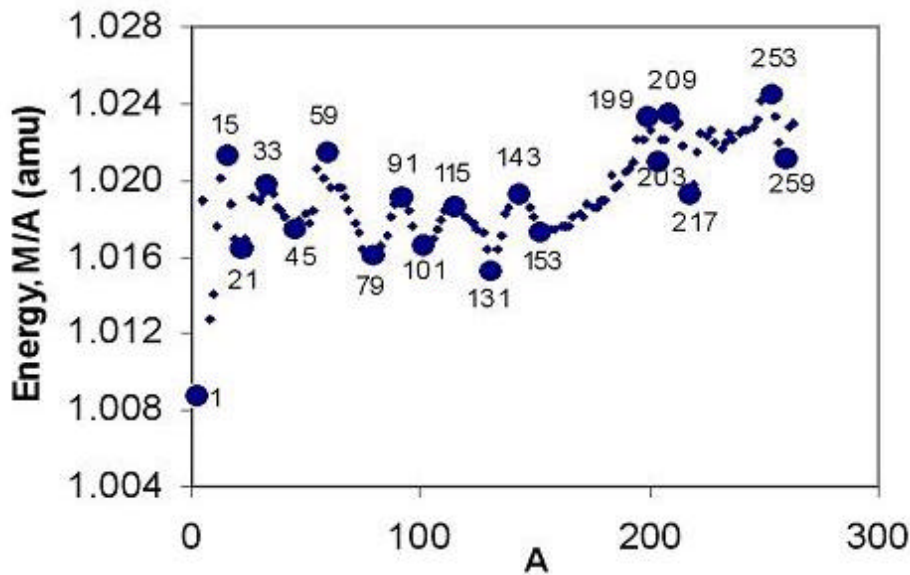

**Figure 5.** Values of M/A at Z/A = 0 for all odd-A parabolas, A = 1-263.

The data in Fig. 5 includes, for example, M/A = 1.019 at A = 27, as shown earlier in Fig. 3. Note that all values of M/A for A > 1 are higher than that of the free neutron at A = 1. This was first recognized as an indication of repulsive interactions between neutrons [17]. Neutron emission from these nuclides would typically generate about 10 MeV per nucleon, as shown by the example in Fig. 3 at A = 27.

The rhythmic distribution with A in values of M/A at Z/A = 0 was not understood in 2000. However, the peaks and valleys in Fig. 5 occur at the same mass numbers as those in Fig. 4 for A ≥ 79. Nuclear clustering into tightly packed structures produces peaks at A = 91, 115, 143, 199, 209 and 253 in Fig. 4 from enhanced Coulomb repulsion. Nuclear clustering into tightly packed structures produce peaks at these same mass numbers in Fig. 5 from enhanced repulsion between neutrons. Loosely packed nucleons produce valleys at A = 79, 101, 131, 153, 203, 217 and 259 in Fig. 4 from reduced Coulomb repulsion between loosely packed protons and in Fig. 5 from reduced repulsion between loosely packed neutrons.

The rhythmic scatter of data in Fig. 5 suggests that nuclear clustering also occurs below A = 79. However, the positive charge on light nuclei apparently maintains a spherical shape. Thus, the Coulomb energy is proportional to $A^{2/3}$ at A < 79 in Fig. 4, as well as in ordinary mirror nuclides [20].

## 4. Theoretical and experimental studies needed

The structure of the solar core likely involves a central neutron star surrounded by iron-rich material. In order to see if neutron emission from the central neutron star might trigger a series of reactions that generate solar luminosity, neutrinos, and an outpouring of $H^+$ ions from the solar surface [14], a better theoretical understanding is needed of:

a) repulsive interactions between neutrons,
b) clustering of nucleons, and
c) neutron emission by penetration of a gravitational barrier.

Likewise, the proposed structure of the solar core can be tested by experimental measurements to look for evidence of:

d) low energy (E < 0.782 MeV), anti-neutrinos coming from neutron decay near the solar core,
e) another source for the neutral neutrino current detected by SNO experiment [21], and
f) a dense object (about 10 km) at the solar core.

The empirical basis for concluding the likely involvement of processes a) – c) in the operation of the Sun was presented above. However, a better theoretical basis for these processes is needed.

The presence of process d) could be detected by measuring inverse β-decay induced by low energy anti-neutrinos coming from the Sun. For example, the $^{35}Cl \longrightarrow\ ^{35}S$ reaction might produce measurable levels of 87-day $^{35}S$ in the Homestake Mine or in underground deposits of salt (NaCl).

Regarding item e), the SNO experiment [21] on solar neutrinos shows that the charge current comes from the direction of the Sun, but new measurements are needed to determine the source of the much larger neutral current.

A recent paper [15] suggests that the 22-year cycle of solar magnetic storms may arise from the neutron star at the solar core and/or from the iron-rich super-conducting material that surrounds it. Measurements of gravity anomalies and of the Sun's quadrupole moment might also provide information on f), a dense object at the solar core.

## Acknowledgements

This study was supported by the University of Missouri-Rolla and the Foundation for Chemical Research, Inc., which kindly consented to our request to reproduce these figures from earlier reports to the Foundation for Chemical Research, Inc. This report was prepared for presentation at the Fourth, BEYOND 03 Conference, at Castle Ringberg, Germany on 9-14 June 2003. Parts were later presented as a plenary lecture at the 6th Workshop on "Quantum Field Theory Under the Influence of External Conditions" (QFEXT03) at the University of Oklahoma, Norman, 15-19 September 2003.